\UseRawInputEncoding
\pdfoutput=1
\documentclass[journal,twoside,web]{ieeecolor}
\usepackage{tmi}
\usepackage{cite}
\usepackage{amsmath,amssymb,amsfonts}
\usepackage{multirow}
\usepackage{algorithm}
\usepackage{algpseudocode}
\usepackage{algpseudocode,algorithm} 
\usepackage{graphicx}
\usepackage{textcomp}
\usepackage{makecell}
\usepackage{mathrsfs}
\usepackage{wrapfig}
\usepackage{colortbl}
\usepackage[utf8]{inputenc} 
\usepackage[T1]{fontenc}    
\usepackage{url}            
\usepackage{amsfonts}       
\usepackage{nicefrac}       
\usepackage{microtype}      
\usepackage{multirow}
\usepackage{booktabs}

\usepackage[table,xcdraw]{}
\usepackage{tabularx}
\usepackage{hyperref}

\algnewcommand{\LineComment}[1]{\State \(\triangleright\) #1}

\def\BibTeX{{\rm B\kern-.05em{\sc i\kern-.025em b}\kern-.08em
    T\kern-.1667em\lower.7ex\hbox{E}\kern-.125emX}}
\begin{document}
\title{JSMoCo: Joint Coil Sensitivity and Motion Correction in Parallel MRI with a Self-Calibrating Score-Based Diffusion Model}
\author{Lixuan Chen,
Xuanyu Tian, 
Jiangjie Wu,
Ruimin Feng, 
Guoyan Lao,
Yuyao Zhang,
Hongjiang Wei
\thanks{Corresponding author: Hongjiang Wei.}
\thanks{L. Chen, X. Tian, J. Wu, Y. Zhang are with the School of Information Science and Technology, ShanghaiTech University, Shanghai, China.}
\thanks{R. Feng, G. Lao, H. Wei are with the School of Biomedical Engineering, Shanghai Jiao Tong University, Shanghai, China. The National Engineering Research Center of Advanced Magnetic Resonance Technologies for Diagnosis and Therapy (NERC-AMRT), Shanghai Jiao Tong University, Shanghai, China. (e-mail: hongjiang.wei@sjtu.edu.cn).}}

\maketitle

\begin{abstract}
Magnetic Resonance Imaging (MRI) stands as a powerful modality in clinical diagnosis. However, it is known that MRI faces challenges such as long acquisition time and vulnerability to motion-induced artifacts. 
Despite the success of many existing motion correction algorithms, there has been limited research focused on correcting motion artifacts on the estimated coil sensitivity maps for fast MRI reconstruction.
Existing methods might suffer from severe performance degradation due to error propagation resulting from the inaccurate coil sensitivity maps estimation.
In this work, we propose to jointly estimate the motion parameters and coil sensitivity maps for under-sampled MRI reconstruction, referred to as JSMoCo. However, joint estimation of motion parameters and coil sensitivities results in a highly ill-posed inverse problem due to an increased number of unknowns. To address this, we introduce score-based diffusion models as powerful priors and leverage the MRI physical principles to efficiently constrain the solution space for this optimization problem. Specifically, we parameterize the rigid motion as three trainable variables and model coil sensitivity maps as polynomial functions. 
Leveraging the physical knowledge, we then employ Gibbs sampler for joint estimation, ensuring system consistency between sensitivity maps and desired images, avoiding error propagation from pre-estimated sensitivity maps to the reconstructed images.
We conduct comprehensive experiments to evaluate the performance of JSMoCo on the fastMRI dataset. The results show that our method is capable of reconstructing high-quality MRI images from sparsely-sampled k-space data, even affected by motion. It achieves this by accurately estimating both motion parameters and coil sensitivities, effectively mitigating motion-related challenges during MRI reconstruction.
\end{abstract}


\section{Introduction}
\label{sec:introduction}
\IEEEPARstart{M}agnetic Resonance Imaging (MRI) is a leading modality in both clinical diagnosis and fundamental research. 
Nevertheless, one major drawback of MRI is its long data acquisition time. Various acceleration techniques have been developed to reconstruct high-quality MRI images from partially-sampled k-space data. 
Parallel imaging utilizes the redundancy of information between multiple receiver coils to achieve acceleration, and it is widely incorporated into MRI scanners for routine clinical scans \cite{GRAPPA, SENSE}. 
Additionally, the compressed sensing theory provides a more efficient approach for accelerating MRI. It requires a randomly under-sampled k-space, and the resulting incoherent artifacts can be alleviated by applying sparse constraints in the transform domain for the reconstruction \cite{Sparse-MRI, CS-MRI-cardiac}. 
In recent years, deep learning has emerged as a formidable tool for accelerating MRI. These methods train neural networks in a supervised or self-supervised manner, leading to remarkable achievements \cite{DL-MRI-MoDL, DL-MRI-Self-super}. 
Despite the efforts to accelerate MRI, it is still susceptible to patient motion during the data acquisition, which will lead to various types of artifacts and reduce the quality of the reconstructed images~\cite{MoCo-Review}. 
This decline in image quality could result in non-diagnostic information, potentially necessitating the need for a rescan. This might also subsequently lead to treatment delays and increased medical costs. Furthermore, if issues arising from motion are not detected, there is a risk of encountering false positive or negative results \cite{Motion-Quantify}.


Previously, the challenge of motion correction has been widely addressed through two primary approaches: \textit{prospective} and \textit{retrospective} strategies. 
Prospective motion correction strategies involve adapting the acquisition process to compensate for measured rigid-body motion.
However, these methods often require modification to the pulse sequences using navigators or extra detectors. The modification can add complexity to the scan processing, potentially leading to longer scan time.
Alternatively, retrospective methods address motion correction algorithmically after data acquisition, eliminating the need for external hardware modifications.
For retrospective techniques, there are two lines of work.
The first~\cite{usman2020retrospective, armanious2020unsupervised, johnson2019conditional} approach treats motion correction as an image post-processing problem (\textit{i.e} deblurring), 
neglecting the physical forward model of MRI acquisition.
The second line of work incorporates \textit{prior} motion information into a physical model, which accounts for the effect of patient motion on the k-space data. 
The model-based methods alternatively or jointly optimize the image and motion parameters by maximizing the data consistency~\cite{8252880, cordero2018three, batchelor2005matrix}. 
Given the non-convex and highly ill-posed nature of this reconstruction problem, model-based methods struggle to provide the stable performance required for clinical application~\cite{hossbach2023deep}.


Nowadays, score-based diffusion models have emerged to provide powerful deep generative priors for inverse problems. 
Several works have leveraged diffusion models as priors through posterior sampling for solving inverse problems, demonstrating remarkable potential across various tasks (\textit{e.g.} image inpainting, image super-resolution~\cite{CCDF, DPS} and medical imaging~\cite{SDE-Inverse, Robust, Chung-MRI, levac2023accelerated, Ye-motion, Optim-Pattern}). 
For accelerating MRI, several methods based on diffusion models have achieved notable progress~\cite{Chung-MRI, Robust, peng2023one, yu2023universal, cui2023spirit, cao2022high, cui2022self}. 
However, the use of score-based diffusion models to solve the motion correction challenge in under-sampled MRI acquisition has not undergone thorough exploration.
Recently, Levac \textit{et al.} \cite{levac2023accelerated} proposes to jointly reconstruct under-sampled MRI data and estimate motion parameters using score-based diffusion models, achieving state-of-the-art (SOTA) performance in motion correction.

However, both model-based and score-based methods have overlooked the effect of motion on the estimated coil sensitivity maps, which are a key component of multi-coil under-sampled MRI reconstruction. Previous approaches have typically assumed that the coil sensitivity maps are pre-determined without motion or the motion does not affect the fully sampled central portion of the k-space used for calibration. 
However, any motion occurring during the scan for collecting the autocalibration k-space region can potentially introduce motion artifacts, thereby degrading the quality of the reconstructed images.

In this paper, we propose a self-calibrating method jointly estimating the motion parameters and coil sensitivity maps (JSMoCo) for accelerated MRI reconstruction. 
This joint optimization approach ensures system consistency and avoids errors caused by pre-estimated sensitivity maps from the motion-corrupted k-space, resulting in high-quality MRI images free from noticeable artifacts.
Since the joint optimization increases the number of unknowns, rendering the inverse problem more ill-conditioned. 
To address this challenge, we introduce score-based diffusion models as powerful priors and leverage the physical acquisition process in multi-coil MRI to efficiently constrain the solution space of the optimization problem. Specifically, we parameterize the 2D motion parameters using three trainable variables (a rotation angle and two translation offsets), enabling accurate modeling of rigid motion. 
Additionally, we parameterize the coil sensitivity maps using polynomial functions, explicitly enforcing the continuous and smooth characteristics of coil sensitivity maps. 
Leveraging this physical knowledge, we then employ the Gibbs sampler~\cite{Gibbs} to jointly optimize the motion parameters, coil sensitivity maps, and reconstructed MRI images through sampling from the joint posterior distribution.

To evaluate the effectiveness of the proposed method, we conduct experiments on the fastMRI dataset \cite{knoll2020fastmri}. 
Qualitative and quantitative results demonstrate that JSMoCo yields reconstructions with fewer artifacts under four different levels of rigid motion and three different acceleration rates.
The main contributions of this work are summarized:
\begin{enumerate}
    \item JSMoCo is the first method that considers the effect of motion on the coil sensitivity maps, to the best of our knowledge. 
    Through the joint optimization strategy, it effectively mitigates error propagation from the pre-estimated sensitivity maps to the final reconstructed image, ultimately leading to the recovery of high-quality images with reduced artifacts.

    \item JSMoCo incorporates the multi-coil MRI acquisition process with rigid motion into the score-based diffusion priors, thereby effectively constraining the solution space and efficiently optimizing the highly ill-posed inverse problem.

    \item Through comprehensive experiments, our results demonstrate that the proposed JSMoCo method excels in producing high-quality reconstructions. Notably, it also successfully estimates motion parameters and coil sensitivity maps in a manner consistent with the MRI forward imaging process.
\end{enumerate}

\section{Preliminaries}

\begin{table}
\caption{Meaning of mathematical notations in background and methods.}
\label{notation}
\centering
\begin{tabular}{cl} 
\toprule
\textbf{Notation} & \textbf{Definition}  \\ 
\midrule
   $\mathbf{x}_0$ & Reconstructed MRI image           \\
   $\mathbf{y}$ & The k-space measurements by $c$ coils, $\mathbf{y}=\left\{ \mathbf{y}_i \right\}_{i=1}^{c}$\\
   $\mathbf{P}$ & Undersampling operator \\
   $\mathbf{F}$ & Fourier transform matrix \\
   $\mathbf{S}_{i}$ & Sensitivity map matrix of the $i$-th coil \\
   $\theta_{j}$ & Rotation angle of the $j$-th shot, $\boldsymbol{\theta}=\left\{ \mathbf{\theta}_j \right\}_{j=1}^{J}$\\
   $\mathbf{t}_{j}$ & Translation vector of the $j$-th shot, $\mathbf{t}=\left\{ \mathbf{t}_j \right\}_{j=1}^{J}$\\
   $\mathbf{m}$ & Motion parameters to be estimated, $\mathbf{m}=\{\boldsymbol{\theta}, \mathbf{t}\}$\\
   $\boldsymbol{\varphi}$ & Unknown vector of the CSMs to be estimated\\

\bottomrule
\end{tabular}
\end{table}

\subsection{Multi-coil MRI Reconstruction}
The measurement process of multi-coil MRI can be written as
\begin{equation}
    \mathbf{y}=\mathbf{A} \mathbf{x}_0+\mathbf{n},
    \label{eq:MRI}
\end{equation}
where $\mathbf{x}_0 \in \mathbb{C}^{n}$ represents the image to be reconstructed, $\mathbf{y} =\left\{ \mathbf{y}_i \right\}_{i=1}^{c}$ is the measurements (\textit{i.e.}, k-space signal) from total $c$ receiver coils, and $\mathbf{n}$ is the noise. For each coil channel $i$, the acquired measurements $\mathbf{y}_i \in \mathbb{C}^{m}$ can be expressed as
\begin{equation}
    \mathbf{y}_{i}=\mathbf{P F S}_{i}\mathbf{x}_0+\mathbf{n}_i,
    \label{eq:Multi-MRI}
\end{equation}
where $\mathbf{S}_{i} \in \mathbb{C}^{n \times n}$ denotes the diagonalized sensitivity map matrix of the $i$-th coil, $\mathbf{F} \in \mathbb{C}^{n \times n}$ denotes the Fourier transform matrix, and $\mathbf{P} \in \mathbb{C}^{m \times n}$ is the undersampling operator.

Multi-coil MRI reconstruction is to recover the unknown image $\mathbf{x}_0$ from its undersampled k-space signal $\mathbf{y}$. Due to the incomplete measurements (\textit{i.e.}, $m\ll n$) caused by the under-sampling operator for accelerating acquisition, the inverse problem in MRI reconstruction is ill-posed. For the multi-coil MRI, even if the number of coils is large and $mc>n$, the problem may still be highly ill-conditioned since the coil sensitivity maps have spatial correlations resulting in a linear dependence among equations.

Due to the ill-posed nature of the above problem, prior knowledge about the reconstructed image $\mathbf{x}_0$ is generally imposed in the form of the regularization term to narrow the solution space. 
Thus, it is critical to construct an effective prior that accurately represents the underlying data distribution.
Many approaches for MRI reconstruction rely on sparsity-based~\cite{Sparse-MRI, CS-MRI-cardiac, CS-MRI-kt-SLR} or low-rank~\cite{CS-MRI-Low-Rank, CS-MRI-Low-Rank3} priors. However, the hand-crafted priors often struggle to accurately represent the complex data distribution $p_\textrm{data}$ of MRI scans, potentially limiting the quality of reconstructed images.

\subsection{Score-Based Diffusion Models for MRI Reconstruction}
Score-based diffusion models have recently demonstrated their efficacy as excellent generative priors for solving inverse problems (\textit{e.g.}, super-resolution~\cite{CCDF, DPS}, computed tomography (CT)~\cite{SDE-Inverse, MC-CT}, and compressed-sensing MRI (CS-MRI)~\cite{Robust, SDE-Inverse, Chung-MRI, levac2023accelerated, Ye-motion, Adaptive-MRI, Optim-Pattern}). 
Score-based diffusion models sample the desired prior distribution from a Gaussian distribution by learning the reverse diffusion process. 
Given a diffusion model trained with a large amount of data, we can generate image samples that are consistent with the observed measurements by incorporating the relevant physical forward model.

We briefly revisit the basic fundamental principles of score-based diffusion models.
The forward process diffuses the data distribution into a fixed prior distribution and can be modeled as the solution of the following stochastic differential equation (SDE):
\begin{equation}
    \mathrm{d} \mathbf{x}=\mathbf{f}(\mathbf{x}, t) \mathrm{d} t+g(t) \mathrm{d} \mathbf{w},
    \label{eq:sde-forward}
\end{equation}
where $\mathbf{f}(\mathbf{x}, t)$ and $g(t)$ are the drift and diffusion coefficient respectively, and $\mathbf{w}$ is the standard Wiener process.


The goal of the score-based diffusion model is to generate a data sample $\mathbf{x}_{0} \sim p_{0}=p_\textrm{data}$ by starting from a noise sample $\mathbf{x}_{T} \sim p_{T}$, which can be achieved by the corresponding reverse SDE of Eq. (\ref{eq:sde-forward}):
\begin{equation}
\mathrm{d} \mathbf{x}_t=\left[\mathbf{f}(\mathbf{x}_t, t)-g(t)^{2} \nabla_{\mathbf{x}_{t}} \log p_{t}\left(\mathbf{x}_{t}\right)\right] \mathrm{d} t+g(t) \mathrm{d} \overline{\mathbf{w}},
    \label{eq:sde-reverse}
\end{equation} 
where $\nabla_{\mathbf{x}_{t}} \log p_{t}\left(\mathbf{x}_{t}\right)$ is known as the score function of $p_{t}\left(\mathbf{x}_{t}\right)$, typically approximated by a neural network $\mathbf{s}_{\theta}$ trained with denoising score matching (DSM)~\cite{DSM}:
\begin{equation}
    \theta^{*}=\operatorname{argmin} \mathbb{E}_{t, \mathbf{x}_{t}, \mathbf{x}_{0}}\left[\left\|\mathbf{s}_{\theta}\left(\mathbf{x}_{t}, t\right)-\nabla_{\mathbf{x}_{t}} \log p\left(\mathbf{x}_{t} | \mathbf{x}_{0}\right)\right\|_{2}^{2}\right]. 
    \label{eq:DSM}
\end{equation}

Consider the MRI reconstruction problem in Eq. (\ref{eq:MRI}), the goal is to sample from the \textit{posterior} distribution $p(\mathbf{x}_0|\mathbf{y})$. Utilizing the Bayes’ rule $p(\mathbf{x}_0|\mathbf{y})=p(\mathbf{x}_0) p(\mathbf{y}|\mathbf{x}_0) / p(\mathbf{y})$ and leveraging the pre-trained diffusion model as the \textit{prior}, it is straightforward to modify Eq. (\ref{eq:sde-reverse}) to arrive at the reverse diffusion sampler for sampling from the posterior distribution: 
\begin{equation}
\begin{aligned}
    \mathrm{d} \mathbf{x}_t=[\mathbf{f}(\mathbf{x}_t, t)-g(t)^{2} & (\nabla_{\mathbf{x}_{t}} \log p_{t}(\mathbf{x}_{t})\\
     + & \nabla_{\mathbf{x}_{t}} \log p_{t}(\mathbf{y} | \mathbf{x}_{t}))] \mathrm{d} t+g(t) \mathrm{d} \overline{\mathbf{w}}.
    \label{eq:inverse-problem}
\end{aligned}
\end{equation}
To compute Eq. (\ref{eq:inverse-problem}), it is necessary to obtain the score function $\nabla_{\mathbf{x}_{t}} \log p_{t}(\mathbf{x}_{t})$ and the likelihood $\nabla_{\mathbf{x}_{t}} \log p_{t}(\mathbf{y}| \mathbf{x}_{t})$. The score function term can be obtained by the pre-trained score function $\mathbf{s}_{{\theta}^*}$. However, obtaining the likelihood term in a closed-form is challenging because it depends on the time $t$.

Since the noise $\mathbf{n}$ in Eq. (\ref{eq:MRI}) is assumed to be Gaussian noise with variance $\sigma^2$, then $p\left(\mathbf{y}| \mathbf{x}_{0}\right) \sim \mathcal{N}\left(\mathbf{y} | \mathbf{A}\mathbf{x}_{0}, \sigma^{2} \mathbf{I}\right)$. Jalal \textit{et al.}~\cite{Robust} proposed to approximate the likelihood by introducing a heuristic term $\gamma_{t}$ of assuming higher levels of noise as $t \rightarrow T$ to counteract the incorrectness in estimation. Therefore, the likelihood can be approximated as
\begin{equation}
    \nabla_{\mathbf{x}_{t}} \log p\left(\mathbf{y} \mid \mathbf{x}_{t}\right)\simeq\frac{\mathbf{A}^{H}\left(\mathbf{y}-\mathbf{A}\mathbf{x}_{t}\right)}{\gamma_{t}^{2}+\sigma^{2}}.
    \label{eq:poster-approx}
\end{equation}


\section{Methods}
\begin{figure*}[!t]
	\centerline{\includegraphics[width=0.85\textwidth]{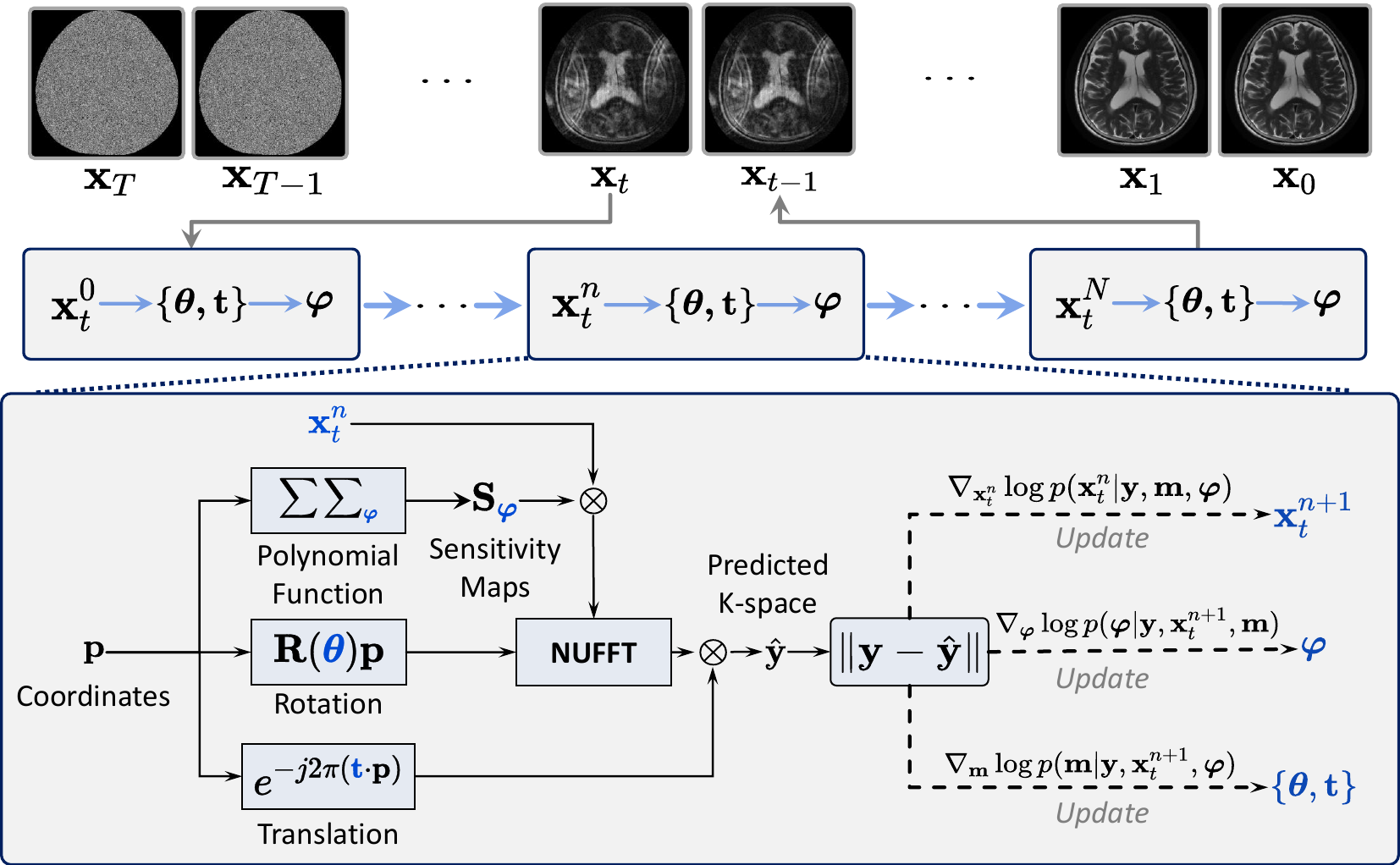}}
	\caption{Overview of the proposed JSMoCo. During the reversing diffusion process (\textit{i.e.}, $t=T\rightarrow 0$), we iteratively sample $\mathbf{x}_t$, $\mathbf{m}$ and $\boldsymbol{\varphi}$ for a total of $N$ times at each timestep $t$.
 }
	\label{fig:csm_result}
\end{figure*}
Although existing retrospective motion correction methods~\cite{8252880, haskell2019network, usman2020retrospective, levac2023accelerated} have shown great performance, a fundamental limitation persists: requiring \textit{accurate} coil sensitivity maps for motion estimation and image reconstruction.
However, inevitable relative motion during the MRI acquisition always results in the \textit{inaccurate} estimation of coil sensitivity maps. 
In this work, we assume the head motion is rigid and model the sensitivity maps as polynomial functions~\cite{JSENSE}.
We then undertake joint optimization of the image and the unknown parameters linked to both motion and sensitivity maps. 
This approach effectively mitigates the propagation of errors from the sensitivity maps to the final reconstructed image.
Consequently, our proposed method can reconstruct high-quality MRI images, even when motion corruption is present in the acquired measurements, especially for coil sensitivity estimation, \textit{e.g.}, the k-space autocalibration regions. 


\subsection{Rigid Motion Parameterization}\label{Motion}
In the proposed model, we make two basic assumptions: 
(1) We limit the type of motion to intra-slice, rigid-body motion while neglecting minor deformable motion (\textit{e.g.}, brain pulsation)~\cite{MoCo-Review, 8252880, haskell2019network}.
(2) We assume that the motion is quasi-static, meaning that objects remain stationary within the same repetition time (TR). This assumption holds merit because the acquisition time of each individual shot is rapid (\textit{e.g.}, in common sequences like fast spin echo (FSE), the intervals between different TRs are typically on the order of seconds~\cite{FSE-Motion, levac2023accelerated}). 

Based on these two assumptions, we define a rotation matrix $\mathbf{R}(\theta_{j})\in \mathbb{SO}(2)$ ($\theta_{j}$ denotes the rotation angle of k-space) and a translation vector $\mathbf{t}_{j} \in \mathbb{R}^2$ to parameterize the rigid motion of the $j$-th item in total $J$ shots:
\begin{equation}
    \mathbf{R}\left(\theta_{j}\right)=\left[\begin{array}{rr}
    \cos \theta_{j} & -\sin \theta_{j} \\
    \sin \theta_{j} & \cos \theta_{j}
    \end{array}\right], \quad \mathbf{t}_{j}=\left[\begin{array}{ll}
    t_{x}^{j} & t_{y}^{j}
    \end{array}\right]^{\top}.
    \label{eq:forward-motion}
\end{equation}

We first perform the rotation operator on the spatial frequency coordinates $\mathbf{p}=(k_{x},k_{y})$, which are related to the readout and phase-encoding directions:
\begin{equation}
    \mathbf{p}_{\boldsymbol{\theta}} = \mathbf{R}\left(\boldsymbol{\theta}\right)\mathbf{p}.
    \label{eq:rotate}
\end{equation}
The rotation operator destroys the uniformity of the equispaced periodic frequencies and requires a Non-Uniform Fourier Transform (NUFFT) to transform the image domain to the frequency domain:
\begin{equation}
    \mathbf{y}_{\boldsymbol{\theta}} = \textrm{NUFFT}\left\{\mathbf{S}_{i}\mathbf{x}_0, \mathbf{p}_{\boldsymbol{\theta}}\right\},
    \label{eq:rotate-forward}
\end{equation}
where $\mathbf{S}_{i}$ denotes the sensitivity map
matrix of the $i$-th coil and $\mathbf{x}_0$ is the motion-free image.

According to the Fourier theorem, object translation in the image domain causes a linear phase in the k-space in the direction of motion~\cite{Motion-form1, Motion-form2}. The motion-corrupted measurement $\mathbf{y}_{\boldsymbol{\theta},\mathbf{t}}$ after performing the translation operation can be expressed as:
\begin{equation}
    \mathbf{y}_{\boldsymbol{\theta},\mathbf{t}} = \mathbf{y}_{\boldsymbol{\theta}} \cdot \exp{\left[-j2\pi(\mathbf{t} \cdot \mathbf{p})\right]}.
    \label{eq:motion-forward}
\end{equation}
Our goal is to estimate the rotation angle $\boldsymbol{\theta}$ and the translation~$\mathbf{t}$.




\subsection{Coil Sensitivity Maps Parameterization}\label{csm}
We model the sensitivity map $\mathbf{S}_{i}$ in Eq. (\ref{eq:Multi-MRI}) as a polynomial function of the spatial coordinates~\cite{JSENSE}. Specifically, the value of the $i$-th coil sensitivity at the coordinate $(x,y)$ can be expressed as:
\begin{equation}
    \mathbf{S}_{i}(x, y)=\sum_{p=0}^{N} \sum_{q=0}^{N} \varphi_{p, q, i} x^{p} y^{q},
    \label{eq:sde-joint}
\end{equation}
where $\varphi_{p, q, i}$ is the coefficient of a polynomial.
Thus, the coefficients of order-$N$ polynomial form the unknown vector $\boldsymbol{\varphi}$ to be optimized.

The polynomial representation inherently yields smooth variations that align with the characteristics of the sensitivity maps. Furthermore, the polynomial representation significantly reduces the number of unknowns in the coil sensitivity maps, thereby alleviating the issue of under-determination in the inverse problem.

\subsection{Joint Parameters Estimation and Image Reconstruction}\label{Joint}
Based on the above parameterization of  rigid motion (Sec.~\ref{Motion}) and coil sensitivity maps (Sec.~\ref{csm}), the multi-coil MRI acquisition process corrupted by the in-plane and rigid motion can be formulated as:
\begin{equation}
    \mathbf{y} = \textrm{NUFFT}\left\{\mathbf{S}_{\boldsymbol{\varphi}}\mathbf{x}_0, \mathbf{R}\left(\boldsymbol{\theta}\right)\mathbf{p}\right\} \cdot \exp{\left[-j2\pi(\mathbf{t} \cdot \mathbf{p})\right]}+\mathbf{n},
    \label{eq:total-forward}
\end{equation}
where $\mathbf{p}$ is the k-space coordinates indicating the desired acquisition trajectory and $\mathbf{n}$ denotes the noise. 
The parameterized forward model described above can be concisely expressed as:
\begin{equation}
    \mathbf{y} = \mathcal{A}_{\mathbf{m, \boldsymbol{\varphi}}}\left(\mathbf{x}_0\right)+\mathbf{n},
    \label{eq:total-forward-simple}
\end{equation}
where $\mathcal{A}_{\mathbf{m}, \boldsymbol{\varphi}}$ is the forward physical model parameterized by motion parameters $\mathbf{m}=\{\boldsymbol{\theta}, \mathbf{t}\}$ and polynomial coefficients $\boldsymbol{\varphi}$ of coil sensitivity maps.
Our objective is to jointly reconstruct the target image $\mathbf{x}_0$ and estimate the parameters $\mathbf{m}, \boldsymbol{\varphi}$ in the physical forward model $\mathcal{A}_{\mathbf{m}, \boldsymbol{\varphi}}$ from the partially-acquired, motion-corrupted measurement $\mathbf{y}$. 

In the Bayesian framework, the optimal solution of this task can be achieved by sampling from the joint posterior distribution $p(\mathbf{x}_0, \mathbf{m}, \boldsymbol{\varphi}|\mathbf{y})$.
However, it is intractable to sample from the posterior distribution directly. 
Thus, we adopt the Gibbs sampler into the reverse diffusion process to sample from the corresponding posterior distribution.
Gibbs sampler is a widely used Markov chain Monte Carlo method for sampling the joint distribution of a set of variables~\cite{Gibbs}. 
It samples the joint distribution through an iterative sampling of each individual variable from their respective conditional distributions, conditioned on all the other variables.
With numerous iterative steps, the sampling strategy will converge to the joint distribution~\cite{heckerman2000dependency}.

In our scenario, we can sample the latent variable $\mathbf{x}_t$ and the unknown parameters $\mathbf{m}, \boldsymbol{\varphi}$ from joint distribution conditional on measurement $\mathbf{y}$ within a cycle of reversing perturbation process (\textit{i.e.}, $t=T\rightarrow 0$). At each time $t$, we iteratively sample $\mathbf{x}_t$ from $p(\mathbf{x}_t|\mathbf{y},\mathbf{m}, \boldsymbol{\varphi})$, $\mathbf{m}$ from $p\left(\mathbf{m}|\mathbf{x}_t, \mathbf{y}, \boldsymbol{\varphi}\right)$, and $\boldsymbol{\varphi}$ from $p\left(\boldsymbol{\varphi}| \mathbf{x}_t, \mathbf{y}, \mathbf{m}\right)$. Based on the theory of Gibbs sampler, it will finally converge to sampling parameters from the intractable joint distribution  $p(\mathbf{x}_t, \mathbf{m}, \boldsymbol{\varphi} | \mathbf{y})$.
Next, we will describe the sampling strategies of $\mathbf{x}_t, \mathbf{m}$ and $\boldsymbol{\varphi}$ and the sampling procedures are shown in Algorithm~\ref{algorithm1}. 

\subsubsection{Sampling Image $\mathbf{x}_t$}
At the time step $t$, we 
sample $\mathbf{x}_t$ from the conditional distribution $p(\mathbf{x}_t|\mathbf{y},\mathbf{m}, \boldsymbol{\varphi})$.
Utilizing the Bayes' rule, the score function of $p(\mathbf{x}_t|\mathbf{y},\mathbf{m}, \boldsymbol{\varphi})$ can be decomposed into two terms.
Meanwhile, due to the independence of parameters $\mathbf{x}_t, \mathbf{m}, \boldsymbol{\varphi}$, the score function can be further simplified as:
\begin{equation}
\begin{aligned}
    \nabla_{\mathbf{x}_{t}}\log p(\mathbf{x}_t|\mathbf{y},\mathbf{m},\boldsymbol{\varphi}) =  \nabla_{\mathbf{x}_{t}}&\log p(\mathbf{y}|\mathbf{x}_t,\mathbf{m},\boldsymbol{\varphi})\\
    &+\nabla_{\mathbf{x}_{t}}\log p(\mathbf{x}_t).
    \label{eq:image-score-bayes}
\end{aligned}
\end{equation}
Eq. (\ref{eq:image-score-bayes}) becomes solvable because the gradient of the log likelihood can be analytically approximated.
Specifically, we leverage the physical forward model $\mathcal{A}_{\mathbf{m}, \boldsymbol{\varphi}}$ to approximate the first term $\nabla_{\mathbf{x}_{t}}\log p(\mathbf{y}|\mathbf{x}_t,\mathbf{m},\varphi)$, which is similar to Eq.~(\ref{eq:poster-approx}):
\begin{equation}
    \nabla_{\mathbf{x}_{t}} \log p\left(\mathbf{y}|\mathbf{x}_t,\mathbf{m},\boldsymbol{\varphi}\right)\simeq\frac{\mathcal{A}_{\mathbf{m, \boldsymbol{\varphi}}}^{H}\left(\mathbf{y}-\mathcal{A}_{\mathbf{m}, \boldsymbol{\varphi}}\left(\mathbf{x}_{t}\right)\right)}{\gamma_{t}^{2}+\sigma^{2}}.
    \label{eq:image-approx}
\end{equation}
The second term $\nabla_{\mathbf{x}_{t}}\log p(\mathbf{x}_t)$ in Eq.~(\ref{eq:image-score-bayes}) is the diffusion priors provided with the pre-trained score function $\mathbf{s}_{{\theta}^*}$ in Eq.~(\ref{eq:DSM}).
After obtaining the score of the conditional distribution, we perform Langevin dynamics~\cite{Langevin} to sample $\mathbf{x}_t$ as follows:
\begin{equation}
    \mathbf{x}_{t} \leftarrow \mathbf{x}_{t} + \lambda_{\mathbf{x},t}\nabla_{\mathbf{x}_{t}} \log p(\mathbf{x}_t|\mathbf{y},\mathbf{m},\boldsymbol{\varphi})+\sqrt{2 \lambda_{\mathbf{x},t}}\boldsymbol{z}, \boldsymbol{z}\sim \mathcal{N}(0 ,\mathbf{I}),
    \label{eq:image-Langevin}
\end{equation}
where ${\lambda}_{\mathbf{x},t}$ is the tunable step-size.

\subsubsection{Sampling Motion Parameters $\mathbf{m}$}
At time step $t$, we sample the motion parameters $\mathbf{m}$ from the conditional distribution $p\left(\mathbf{m}|\mathbf{x}_t, \mathbf{y}, \boldsymbol{\varphi}\right)$.
Similar to the sampling strategies for sampling $\mathbf{x}_t$, we can use the Bayes' rule to decompose the score of conditional distribution $\nabla_{\mathbf{m}} \log p\left(\mathbf{m}|\mathbf{x}_t, \mathbf{y}, \boldsymbol{\varphi}\right)$ into two terms and approximate the score function $\nabla_{\mathbf{m}} \log p(\mathbf{y}|\mathbf{x}_t,\mathbf{m},\boldsymbol{\varphi})$ as:
\begin{equation}
\label{eq:csm-Langevin}
\left\{\begin{matrix}
\begin{aligned}
 &\nabla_{\mathbf{m}} \log p\left(\mathbf{m}|\mathbf{x}_t, \mathbf{y}, \boldsymbol{\varphi}\right) = \nabla_{\mathbf{m}} \log p(\mathbf{y}|\mathbf{x}_t,\mathbf{m}, \boldsymbol{\varphi})\\
&\hspace{15em}+\nabla_{\mathbf{m}} \log p(\mathbf{m}),\\
&\nabla_{\mathbf{m}} \log p(\mathbf{y}|\mathbf{x}_t,\mathbf{m},\boldsymbol{\varphi})\simeq -\frac{1}{2\sigma^2}\nabla_{\mathbf{m}}\|\mathbf{y}-\mathcal{A}_{\mathbf{m, \varphi}}\left(\mathbf{x}_{t}\right)\|^2.
\end{aligned}
\end{matrix}\right.
\end{equation}
As for the $\nabla_{\mathbf{m}} \log p(\mathbf{m})$ term, we define $\mathbf{t}$ in Euclidean space $\mathbb{R}^2$ and $\mathbf{\theta}$ in $\mathbb{SO}(2)$ space to constraint the motion as the rigid motion.
Then, we perform Langevin dynamics to sample $\mathbf{m}$ from the above score.

\subsubsection{Sampling Coefficients $\boldsymbol{\varphi}$ of coil sensitivities}
Similar to the sampling of $\mathbf{m}$ and $\mathbf{x}_t$, we perform Langevin dynamics to sample $\boldsymbol{\varphi}$ from the approximated score function of conditional distribution $\nabla_{\boldsymbol{\varphi}}\log p\left(\boldsymbol{\varphi}|\mathbf{x}_t, \mathbf{y}, \mathbf{m}\right)$, which can be expressed as:
\begin{equation}
\label{eq:csm-Langevin}
\begin{aligned}
\nabla_{\boldsymbol{\varphi}}\log p\left(\boldsymbol{\varphi}|\mathbf{x}_t, \mathbf{y}, \mathbf{m}\right) \simeq -\frac{1}{2\sigma^2}\nabla_{\boldsymbol{\varphi}}\left \| \mathbf{y}-\mathcal{A}_{\mathbf{m, \varphi}}\left(\mathbf{x}_{t}\right) \right \| ^2\\
+\nabla_{\boldsymbol{\varphi}}\log p(\boldsymbol{\varphi}),
\end{aligned}
\end{equation}
where $p(\boldsymbol{\varphi})$ is parameterized as a polynomial representation, which introduces a generic and simple prior to ensure the smooth characteristics of the sensitivity map and further constrain the solution space of the ill-posed inverse problem.

At each time step $t$, the Gibbs sampler alternately samples the latent variables $\mathbf{x}_t$ and the unknown parameters $\left\{\mathbf{m},\boldsymbol{\varphi}\right\}$ for $N$ times. When $t=0$, 
the samples converge towards the target joint distribution $p(\mathbf{x}_0, \mathbf{m}, \boldsymbol{\varphi}|\mathbf{y})$. Throughout this process, we can leverage the capacity of the generative model to effectively represent data when sampling the unknown parameters. This approach facilitates accurate parameter estimation by leveraging the physical-driven priors.

\begin{algorithm}[t]
    \caption{Joint Posterior sampling}
    \label{algorithm1}
    \begin{algorithmic}[1] 
    \Require Motion-corrupted, partially-acquired measurement $\mathbf{y}$, hyperparameters $\boldsymbol{\lambda}$.
    \Ensure Reconstructed image $\mathbf{x}_0$, motion parameters $\mathbf{m}$, polynomial coefficients of the sensitivity map $\boldsymbol{\varphi}$.
    \LineComment{\textcolor[rgb]{0.4,0.4,0.4}{\textit{$K$ counts the total number of updates for parameters}}}
    \State $K \leftarrow 0$
    \LineComment{\textcolor[rgb]{0.4,0.4,0.4}{\textit{Initial parameters}}}
    \State $\mathbf{x}_T^0 \sim \mathcal{N}(\mathbf{0}, \mathbf{I})$,$\mathbf{m}^K\sim \mathcal{N}(\mathbf{0}, \sigma_{\mathbf{m}}^2\mathbf{I})$; $\boldsymbol{\varphi}^K\sim \mathcal{N}(\mathbf{0}, \sigma_{\boldsymbol{\varphi}}^2\mathbf{I})$  
    \For{$t = T-1,\dots, 1$}

    \LineComment{Inner Loops for \textit{sampling} parameters}
    \For{$n = 1,\dots, N$}
    \State $\boldsymbol{z}\sim \mathcal{N}(\mathbf{0},\mathbf{I})$, $\boldsymbol{\xi} \sim \mathcal{N}(\mathbf{0},\mathbf{I}) $, $\boldsymbol{\varepsilon} \sim \mathcal{N}(\mathbf{0},\mathbf{I}) $
    \LineComment{\textcolor[rgb]{0.4,0.4,0.4}{\textit{\textit{Sampling}} $\mathbf{x}_t$}}
    \State $s({\mathbf{x}^n_t}) = \nabla_{\mathbf{x}^n_{t}}\log p\left(\mathbf{x}^{n}_{t}|\mathbf{y},\mathbf{m}^K,\boldsymbol{\varphi}^K\right)$
    
    \State $\mathbf{x}^{n+1}_{t}\leftarrow\mathbf{x}^n_{t}+{\lambda}_{\mathbf{x},t} s(\mathbf{x}^n_{t})+\sqrt{2 {\lambda}_{\mathbf{x},t}}\boldsymbol{z}$
    \LineComment{\textcolor[rgb]{0.4,0.4,0.4}{\textit{Sampling} $\mathbf{m}$}}
    \State $s(\mathbf{m}^K) = \nabla_{\mathbf{m}^K}\log p\left(\mathbf{m}^K |\mathbf{y},\mathbf{x}^{n+1}_t,\boldsymbol{\varphi}^K\right)$
    \State $\mathbf{m}^{K+1}\leftarrow \mathbf{m}^K+{\lambda}_{\mathbf{m},t} s(\mathbf{m}^K)+\sqrt{2 {\lambda}_{\mathbf{m},t}} \boldsymbol{\xi}$
    \LineComment{\textcolor[rgb]{0.4,0.4,0.4}{\textit{Sampling} $\boldsymbol{\varphi}$}}
    \State $s(\boldsymbol{\varphi}^K) = \nabla_{\boldsymbol{\varphi}^K}\log p\left(\boldsymbol{\varphi}^K|\mathbf{y},\mathbf{x}^{n+1}_t,\mathbf{m}^{K+1}\right)$
     \State $\boldsymbol{\varphi}^{K+1}\leftarrow\boldsymbol{\varphi}^K+{\lambda}_{\boldsymbol{\varphi},t} s(\boldsymbol{\varphi}^K)+\sqrt{2 {\lambda}_{\boldsymbol{\varphi},t}} \boldsymbol{\varepsilon}$
     \State $K \leftarrow K+1$
    \EndFor
    \State $\mathbf{x}_{t-1}^0 = \mathbf{x}_{t}^N$ \Comment{\textcolor[rgb]{0.4,0.4,0.4}{Compute $\mathbf{x}_{t-1}$}}
    \EndFor
    \end{algorithmic}
\end{algorithm}

\begin{figure*}[!t]
	\centerline{\includegraphics[width=0.9\textwidth]{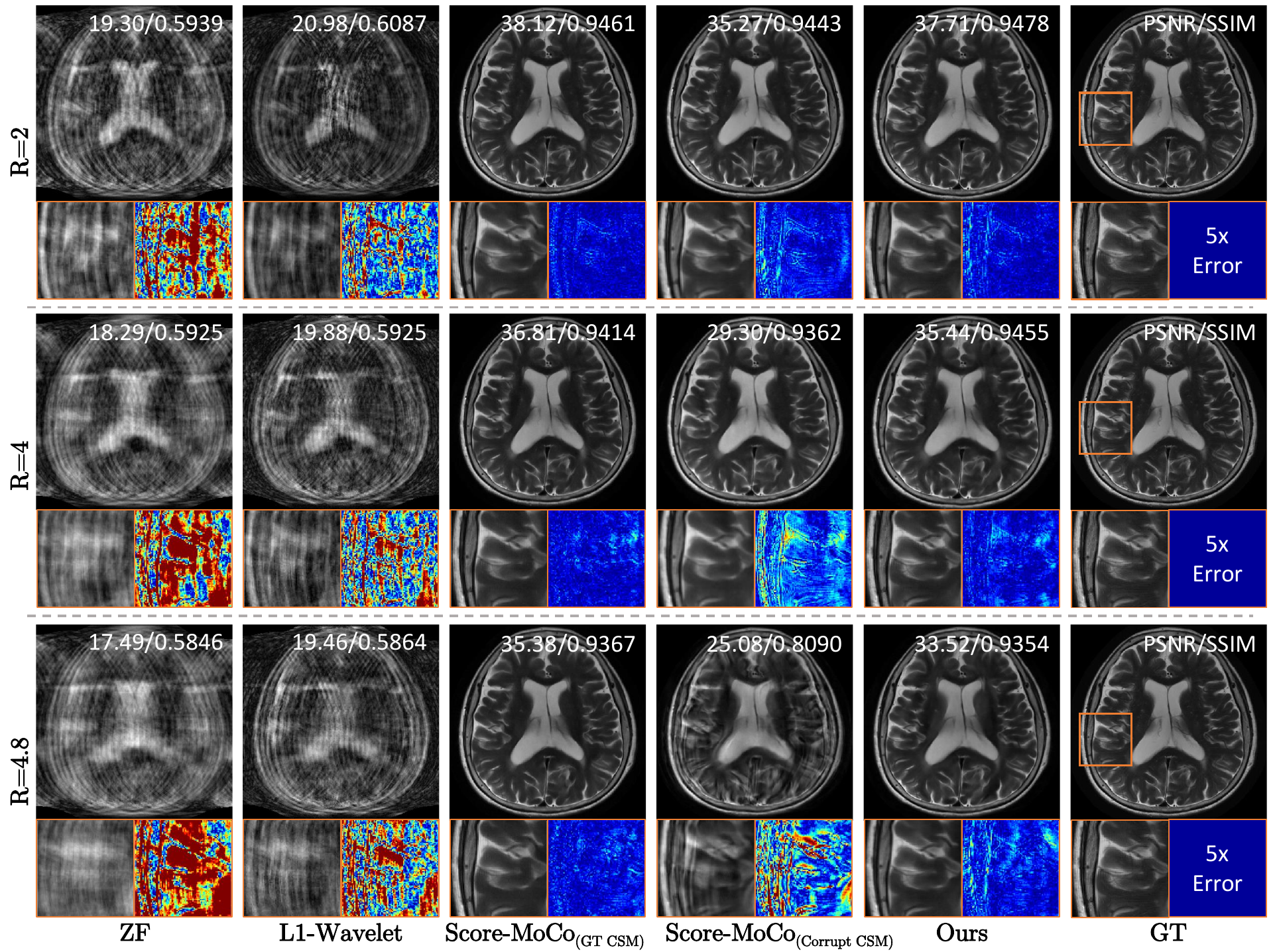}}
	\caption{Qualitative results of the compared methods on the reconstructed images with three accelerated ratios $(R = 2, 4, 4.8)$ that are corrupted by translation of $\pm 3$ pixels and rotation of $\pm 3^\circ$. Quantitative evaluation metrics (PSNR and SSIM) are provided above each image. The $5\times$ error maps are displayed for better visualization.}
	\label{fig:main_result}
\end{figure*}

\subsection{Implementation Details}
For the coil sensitivity maps parameterization, the real and imaginary parts are separately modeled using polynomials (\textit{i.e.}, $|\boldsymbol{\varphi}|=2cN^2$) due to the complex nature of the coil sensitivities. The order of the polynomial function is empirically set to 15.
All the elements in the motion parameters $\mathbf{m}$ and the polynomial coefficients $\boldsymbol{\varphi}$ are initialized with a normal distribution using a constant random seed. 
This is based on the assumption that no prior knowledge of the motion and sensitivities is available.
For the Langevin dynamics, we utilize the \texttt{autograd} implementation in PyTorch to calculate the log density gradient.

In this paper, we focus on developing our method on top of variance exploding (VE)-SDE. In the case of VE-SDE, the drift coefficient $\mathbf{f}(\mathbf{x}, t)=\mathbf{0}$ and the diffusion coefficient $g(t)=\sqrt{\mathrm{d}\left[\sigma^{2}(t)\right]/\mathrm{d} t}$, where $\sigma(t)$ is a monotonically increasing function representing the level of noise added to perturb the data distribution. The network structure of the score function is the same as that of NCSNv2~\cite{NCSNv2}\footnote{https://github.com/ermongroup/ncsnv2}, with the pre-trained score model taken from~\cite{Robust}.

\section{Experiments}

\begin{table*}
\centering
\caption{Quantitative results of compared methods on the fastMRI dataset reconstruction with three accelerated ratios ($R = 2, 4, 4.8$) and four motion parameters settings. Results are presented as PSNR/SSIM. The best and second best results are highlighted in \textbf{bold} and \underline{underline}, respectively.
}
\label{table_image}
\begin{tabular}{cccccccc} 
\toprule
\multirow{2.5}{*}{\begin{tabular}[c]{@{}c@{}}\makecell{\textbf{Accelerate} \\ \textbf{Rate}}\end{tabular}} & \multicolumn{2}{c}{\textbf{Motion Setting}} & \multirow{2.5}{*}{\textbf{ZF}} & \multirow{2.5}{*}{\begin{tabular}[c]{@{}c@{}}\textbf{L1} \\\textbf{Wavelet}\end{tabular}} & \multirow{2.5}{*}{\begin{tabular}[c]{@{}c@{}}\textbf{Score-MoCo}\\\textbf{(GT CSM)}\end{tabular}} & \multirow{2.5}{*}{\begin{tabular}[c]{@{}c@{}}\textbf{Score-MoCo}\\ \textbf{(Corrupt CSM)}\end{tabular}} & \multirow{2.5}{*}{\textbf{Ours}}  \\ 
\cmidrule{2-3}
       & \textbf{Rotation}             & \textbf{Translation}     &                     &                                                                       &                                                                               &                                                                                    &                        \\ 
\midrule
\multirow{4.5}{*}{$\times$2}                                                         & \multirow{2}{*}{$\pm$ 2$^\circ$} & $\pm$ 3 pixels             & 22.28/0.6863        & 24.83/0.7192                                                          & \textbf{35.63}/\textbf{0.9406}                                                                           & 33.49/0.9233                                                                                    & \underline{33.67}/\underline{0.9397}           \\
       &                    & $\pm$ 4 pixels            & 21.44/0.6468        & 22.53/0.6584                                                          &    \textbf{35.10}/\textbf{0.9397}                                                                            &    32.33/0.9202                                                                                 & \underline{33.38}/\underline{0.9378}           \\ 
\cmidrule{2-8}
       & \multirow{2}{*}{$\pm$ 3$^\circ$} & $\pm$ 3 pixels            & 21.94/0.6643        & 23.29/0.6863                                                          &  \textbf{34.87}/\underline{0.9375}                                                                             &   32.92/0.9318                                                                                 & \underline{33.94}/\textbf{0.9391}           \\
       &                    & $\pm$ 4 pixels            & 20.99/0.6292        & 22.83/0.6532                                                          & \textbf{35.30}/\underline{0.9347}                                                                  & 31.12/0.9129                                                                        & \underline{34.65}/\textbf{0.9419}          \\ 
\midrule
\multirow{4.5}{*}{$\times$4}                                                         & \multirow{2}{*}{$\pm$ 2$^\circ$} & $\pm$ 3 pixels            & 19.41/0.6349        & 22.15/0.6471                                                          &   \textbf{33.19}/\textbf{0.9319}                                                                             &     30.19/0.8927                                                                                & \underline{33.13}/\underline{0.9270}           \\
       &                    & $\pm$ 4 pixels            & 19.00/0.6070        & 21.48/0.6152                                                          &         \textbf{34.51}/\textbf{0.9339}                                                                       &  29.79/0.9030                                                                                   & \underline{32.49}/\underline{0.9272}           \\ 
\cmidrule{2-8}
       & \multirow{2}{*}{$\pm$ 3$^\circ$} & $\pm$ 3 pixels            & 19.83/0.6274        & 21.81/0.6312                                                          &   \textbf{33.31}/\textbf{0.9272}                                                       &   30.36/0.9149                                                                                  & \underline{32.47}/\underline{0.9258}           \\
       &                    & $\pm$ 4 pixels            & 19.07/0.5994        & 20.78/0.6010                                                          & \textbf{34.31}/\textbf{0.9256}                                                                  & 31.09/0.8986                                                                        & \underline{32.29}/\underline{0.9233}           \\ 
\midrule
\multirow{4.5}{*}{$\times$4.8}                                                       & \multirow{2}{*}{$\pm$ 2$^\circ$} & $\pm$ 3 pixels            & 19.36/0.6373        & 21.46/0.6501                                                          &       \underline{30.15}/\underline{0.9022}                                                                         &   27.15/0.8493                                                                                  & \textbf{31.95}/\textbf{0.9195}           \\
       &                    & $\pm$ 4 pixels            & 19.07/0.6138        & 20.76/0.6138                                                          &    \textbf{32.74}/\textbf{0.9241}                                                                           &   29.35/0.8703                                                                                 & \underline{30.57}/\underline{0.9136}           \\ 
\cmidrule{2-8}
       & \multirow{2}{*}{$\pm$ 3$^\circ$} & $\pm$ 3 pixels            & 19.43/0.6208        & 22.06/0.6388                                                          &    \textbf{32.44}/\underline{0.9076}                                                                            &   28.75/0.8738                                                                                  & \underline{31.41}/\textbf{0.9180}           \\
       &                    & $\pm$ 4 pixels            & 17.98/0.5892        & 19.68/0.5886                                                          & \underline{31.67}/\textbf{0.9220}                                                                  & 27.93/0.8612                                                                      & \textbf{31.75}/\underline{0.9218}           \\
\bottomrule
\end{tabular}
\end{table*}

\subsection{Dataset and Motion Simulation}
\subsubsection{Dataset}
All the experiments are based on the T2-weighted brain dataset of fastMRI database with approval from the New York University School of Medicine Institutional Review Board~\cite{knoll2020fastmri}. 
We extract 2D slices with the matrix size of $384\times 384$ from 3D volumes as experimental data.
The datasets were acquired using 14, 16 or 20-channel receiver coils.
5 slices from each subject and total 5 subjects were used to evaluate the performance of different approaches.

\subsubsection{Motion Simulation}
We simulate motion-corrupted, undersampled data from the fully-sampled, motion-free k-space data. The root-sum-of-squares reconstruction of the magnitude images from the fully-sampled k-space data provides the ground truth (GT) images.
The ground truth of the coil sensitivity map (CSM) is estimated using ESPIRiT~\cite{uecker2014espirit} based on the fully-sampled, motion-free k-space data.
We synthesize motion-corrupted measurements according to the forward process described in section~\ref{Motion}. More specifically, for each TR interval, the k-space points are translated by $t_x$ and $t_y$ pixels along the X and Y axes, respectively, followed by a rotation of $\theta^\circ$ around the origin. The translation and rotation motion parameters in each shot are sampled from the uniform distribution, denoted as $\mathcal{U}(-k_t, k_t)$ and $\mathcal{U}(-k_{\theta}, k_{\theta})$, respectively. We set $k_t=\{3, 4\}$ and $k_{\theta}=\{2, 3\}$ to simulate four different levels of rigid motion. According to the fully-sampled, motion-corrupted k-space data, we apply the ESPIRiT~\cite{uecker2014espirit} to estimate the motion-corrupted CSM.
In all experiments, we simulate random rigid motion on fully sampled k-space data and then apply them to different scan acceleration ratios ($R=2, 4, 4.8$).

\begin{figure}[!t]
	\centerline{\includegraphics[width=\columnwidth]{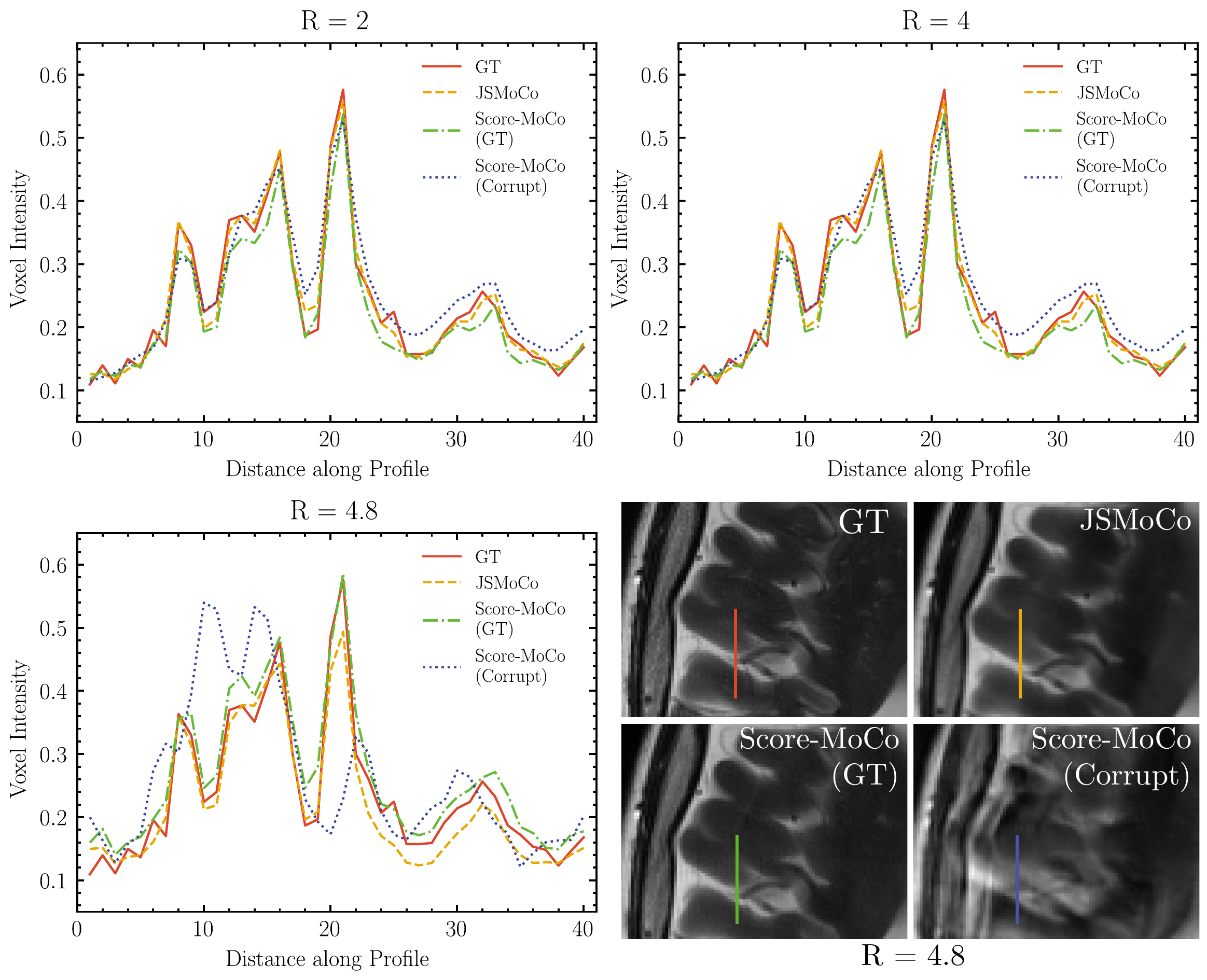}}
	\caption{Comparison of intensity profiles on the reconstructed images obtained using different methods with simulated data corrupted by translation of $\pm 3$ pixels and rotation of $\pm 3^\circ$. }
	\label{fig:profile}
\end{figure}

\subsection{Comparison Methods}
We compared our method with two accelerated MRI reconstruction baselines: 
\subsubsection{Zero-Filling (ZF)}
Zero-filling is a classical method that populates the undersampled k-space regions with zeros.

\subsubsection{L1-wavelet~\cite{lustig2007sparse}}
$\ell_1$-wavelet regularized reconstruction algorithm aims to solve the optimization of the inverse problem with $\ell_1$-sparsity in the
wavelet domain.

\begin{table}
\caption{Quantitative results (NRMSE) of the estimated CSMs using the proposed method, compared with those obtained from motion-corrupted k-space signals using ESPIRiT, with reference to the ground truth CSMs. The best and second best results are highlighted in \textbf{bold} and \underline{underline}, respectively.}
\label{table_csm}
\centering
\begin{tabular}{lcccc} 
\toprule
\textbf{Rotation}       & \multicolumn{2}{c}{$\pm$ $2^\circ$} & \multicolumn{2}{c}{$\pm$ $3^\circ$}  \\
\textbf{Translation }   & $\pm\ 3$ pixels & $\pm\ 4$ pixels          & $\pm\ 3$ pixels & $\pm\ 4$ pixels    \\ 
\midrule
\textbf{Ours ($ R = 2$)}  & \textbf{0.0093}  & \textbf{0.0093} & \textbf{0.0091}   & \textbf{0.0099}           \\
\textbf{Ours ($R = 4$)}   & \underline{0.0096}  & \underline{0.0097} & \underline{0.0105}   & \underline{0.0105}           \\
\textbf{Ours ($R = 4.8$)} & 0.0101  & 0.0111 & 0.0110   & 0.0107           \\
\cmidrule(lr){1-5}
\textbf{Corrputed}     & 0.0228  & 0.0274                 & 0.0233        & 0.0270           \\
\bottomrule
\end{tabular}
\end{table}

\begin{figure}[!t]
	\centerline{\includegraphics[width=\columnwidth]{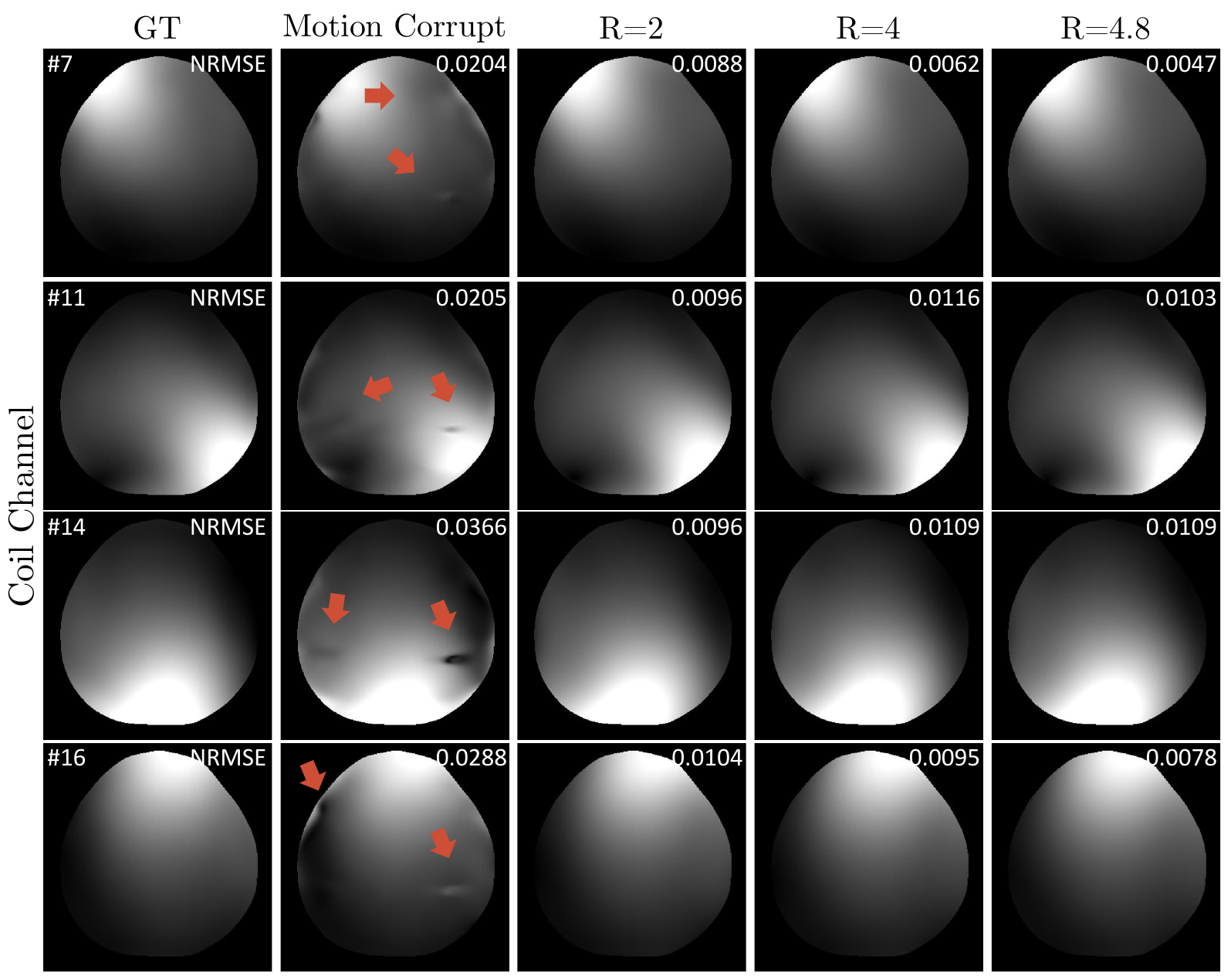}}
	\caption{Qualitative results of estimated CSMs by the proposed method and the ESPIRiT algorithm using motion-corrupted k-space signal. The red arrows point to the artifacts in the CSMs estimated by ESPIRiT. The coil sensitivity maps from four representative channels are presented. }
	\label{fig:estimate_csm_result}
\end{figure}

The above two methods only perform MRI reconstruction without motion correction. We also compared with one score-based method for motion correction in accelerated MRI:
\subsubsection{Score-MoCo~\cite{levac2023accelerated}}
Score-MoCo does not account for the impact of motion on the CSM and relies solely on the CSM estimated from fully-sampled, motion-free k-space data. However, this scenario is somewhat idealistic, as obtaining motion-free k-space data could be challenging in certain scenarios, \textit{e.g.}, for pediatric or Parkinson's disease patients.
In our work, the CSM is jointly estimated during the reconstruction process, producing results consistent with rigid motion.
Thus, we compared Score-MoCo using two different types of estimated CSM:

\begin{enumerate}
    \item Score-MoCo (GT CSM) uses the ground-truth CSM for motion correction. This method is an upper bound on joint estimation performance because it assumes the availability of the ideal CSM derived from motion-free k-space data.

    \item Score-MoCo (Corrupted CSM) uses the CSM estimated from the fully-sampled k-space signal corrupted by motion. This aligns more closely with real-world MRI acquisition scenarios.
\end{enumerate}

\subsection{Evaluation Metrics}
We employ the Peak-Signal-to-Noise Ratio (PSNR) and Structured Similarity Index Measurement (SSIM) to evaluate the quality of the reconstructed MRI images.
For estimated coil sensitivity maps, we compute normalized root-mean-square error (NRMSE) to evaluate the accuracy of estimation.

\section{Results}
\subsection{Results of the Reconstruction Images}
\begin{figure}[!t]
	\centerline{\includegraphics[width=\columnwidth]{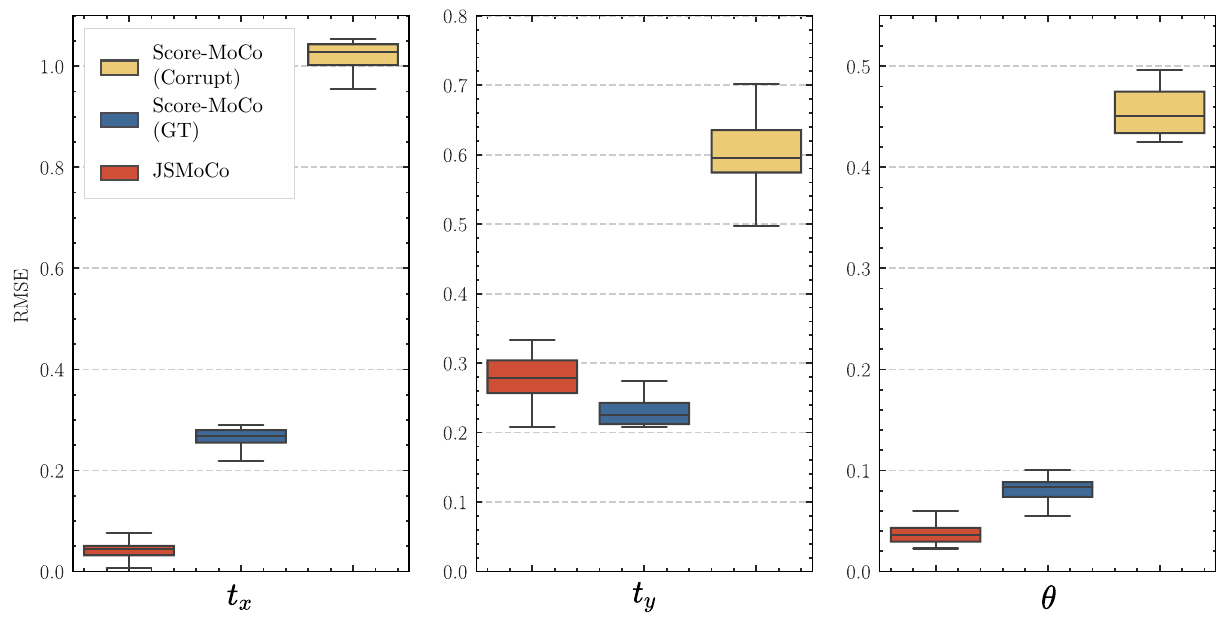}}
	\caption{Comparison of the estimated motion parameters using different methods at $R=4$ corrupted by translation of $\pm 4$ pixels and rotation of $\pm 2^\circ$.}
	\label{fig:motion_result}
\end{figure}

\begin{figure}[!t]
	\centerline{\includegraphics[width=\columnwidth]{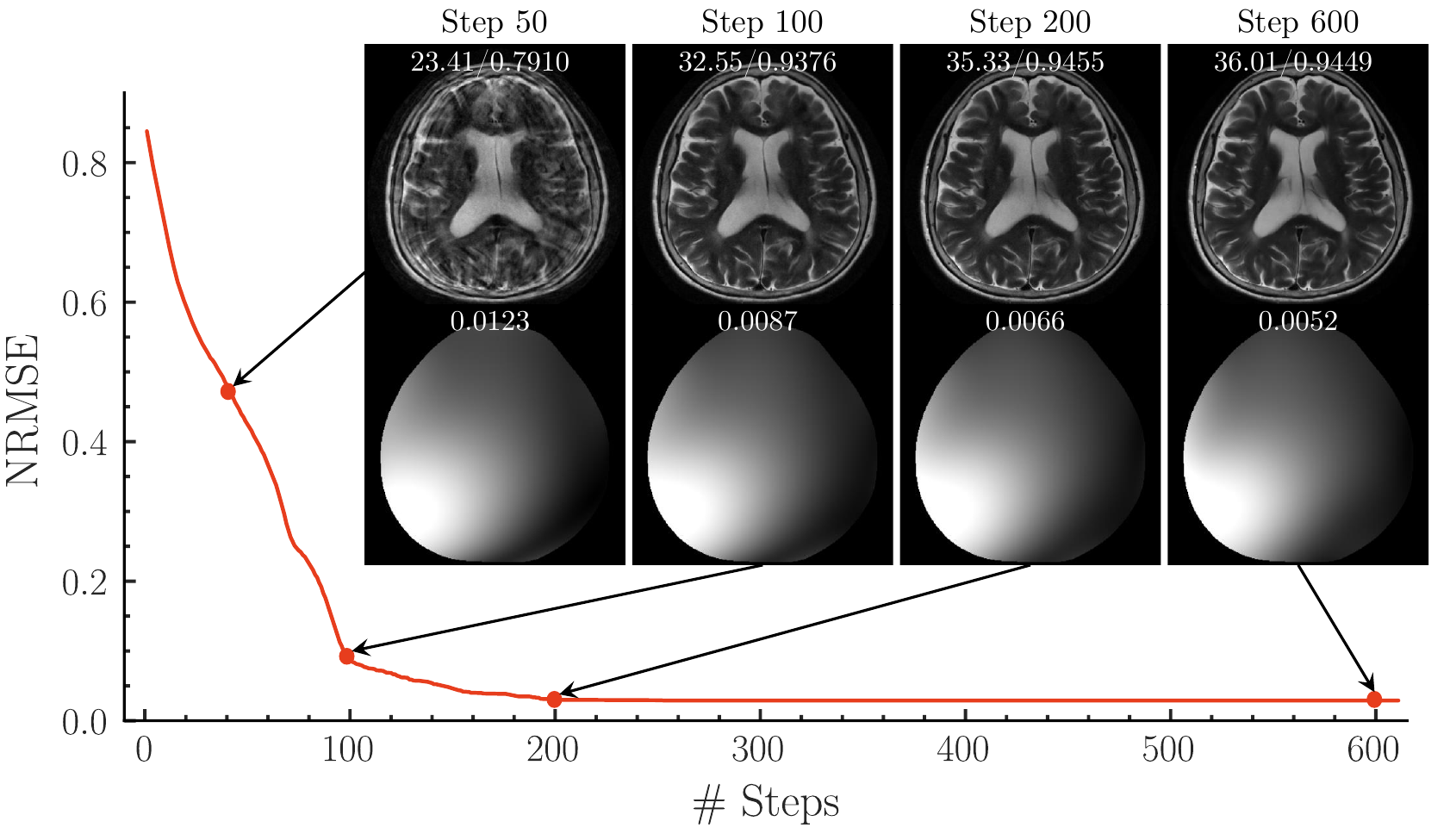}}
	\caption{The convergence of the proposed framework at $R=4$, corrupted by translation of $\pm 3$ pixels and rotation of $\pm 3^\circ$. The red line represents the NRMSE between estimated motion parameters and the ground truth at each step. The top row displays the reconstruction results with PSNR and SSIM values, while the bottom row shows the estimated CSMs with their respective NRMSE values.}
	\label{fig:motion_t}
\end{figure}

\begin{figure}[!t]
	\centerline{\includegraphics[width=\columnwidth]{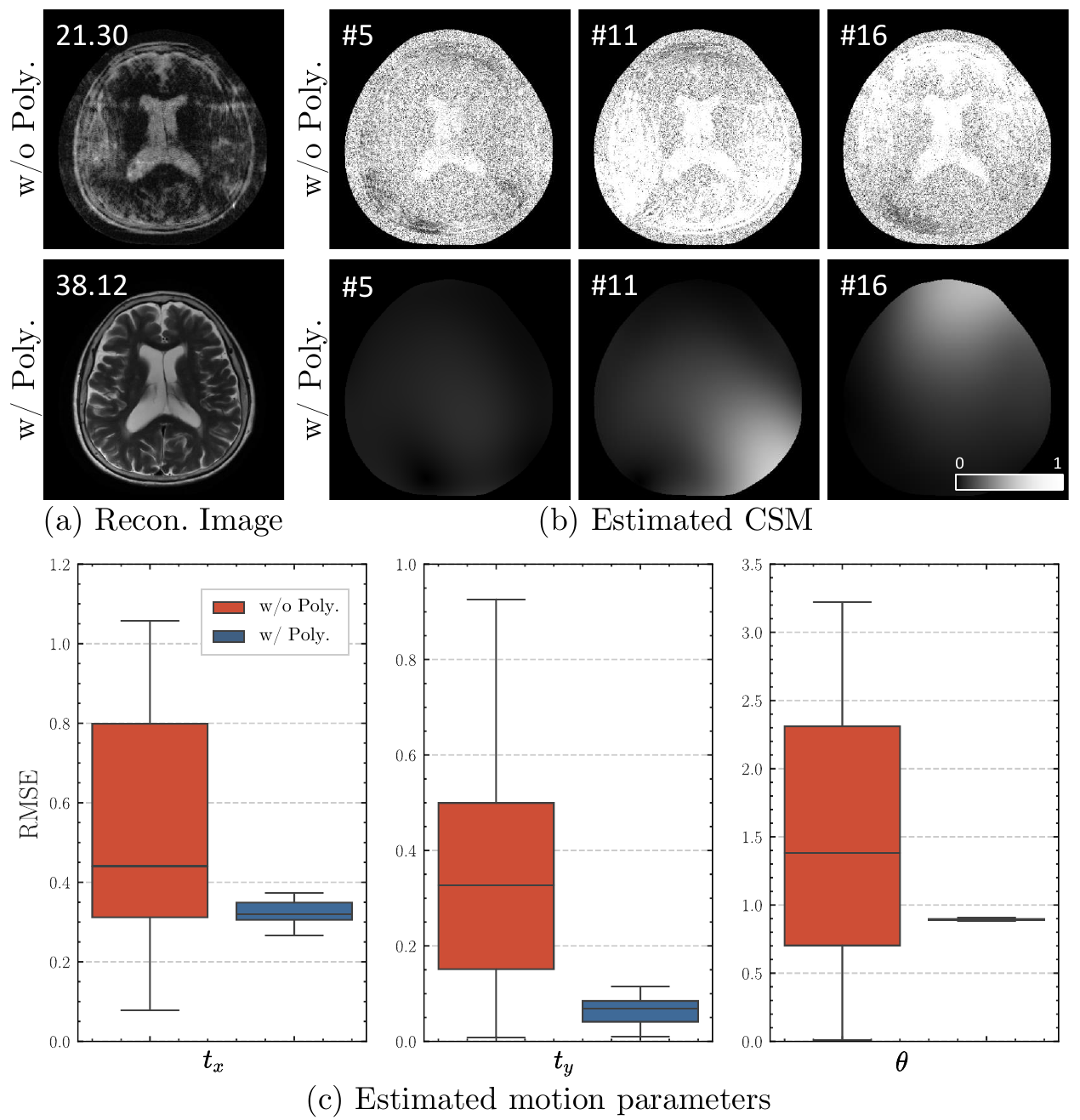}}
	\caption{Qualitative results for the reconstructed images and estimated CSMs, as well as quantitative results for estimated motion parameters with and without modeling CSM as a polynomial function. }
	\label{fig:ablation}
\end{figure}

\begin{figure*}[!t]
	\centerline{\includegraphics[width=\textwidth]{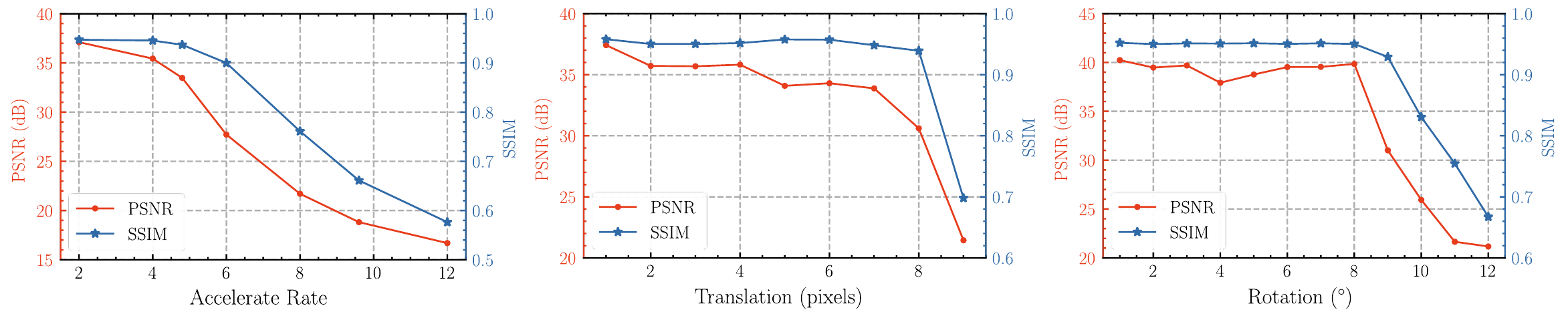}}
	\caption{Reconstruction performance (PSNR and SSIM) with different acceleration rates and different levels of motion.}
	\label{fig:model_capacity}
\end{figure*}

Fig.~\ref{fig:main_result} displays a representative slice reconstructed by different methods with fixed motion offsets and different acceleration factors.
At $R=2$, reconstruction results of ZF and L1-Wavelet suffer from noticeable motion and aliasing artifacts, primarily stemming from the rigid motion corruption. 
In contrast, motion correction with Score-MoCo (corrupted CSM) demonstrates improved reconstruction quality but still struggles with certain suboptimal reconstructions due to the effect of motion on the estimated coil sensitivities.
Our method, which jointly optimizes the sensitivity maps during the reconstruction process, yields faithful MR images. 
It effectively eliminates motion and aliasing artifacts, demonstrating comparable performance to Score-MoCo (with GT CSM), as illustrated in the zoomed-in images in Fig.~\ref{fig:main_result}. 
Similar results are observed when the acceleration factor is raised to 4 and 4.8. Particularly, Score-MoCo with corrupted CSM introduces visible artifacts with a higher acceleration factor. In contrast, our method maintains competitive artifact-reduction results comparable to GT and Score-MoCo (GT CSM). These results demonstrate the robustness of our method in handling higher acceleration factors.


Table~\ref{table_image} reports the quantitative evaluation metrics analyzed on the simulation brain dataset with different acceleration factors and motion settings.
The ZF and L1-Wavelet are vulnerable to corrupted sensitivity maps and motion offsets, producing poor reconstruction results.
Our proposed model outperforms Score-MoCo (corrupted CSM) and achieves competitive reconstruction errors when compared to Score-MoCo (GT CSM) across all motion settings.
For example, our method achieves a PSNR of 33.13 dB, which is more than 3 dB higher than the PSNR achieved by Score-MoCo (corrupted CSM) at $R=4.8$ with the motion offset ($\pm 2^{\circ} \& \pm 3$) pixels.
Notably, at $R=4.8$, JSMoCo demonstrates a comparable reconstruction performance as the Score-MoCo (GT CSM). This is primarily attributed to our joint optimization strategy, which yields a consistent and coherent solution, thus ensuring superior system consistency. 


As shown in Fig.~\ref{fig:profile}, we compare the intensity profiles of the operculum to evaluate the reconstructed local image contrast between tissues. 
The reconstruction results obtained with the proposed JSMoCo and Score-MoCo (GT CSM) exhibit sharp signal changes in the intensity profiles, consistent with the pattern observed on the GT image.
However, in the case of Score-MoCo (corrupted CSM), particularly at $R=4.8$, there is a noticeable shift in the intensity profile when compared to the GT image, indicating the presence of artifacts.

\subsection{Results of the Sensitivity Maps Estimation}
Fig.~\ref{fig:estimate_csm_result} shows the ability of our method to estimate sensitivity maps under different acceleration factors on the simulation brain dataset. Visually, the sensitivity maps estimated by JSMoCo align well with those estimated by ESPIRiT. Conversely, noticeable artifacts are apparent in the CSMs estimated from motion-corrupted k-space data, as indicated by the red arrows on the motion-corrupted maps.
We report the quantitative results in Table~\ref{table_csm}, using the NRMSE calculated relative to the CSMs estimated from fully-sampled, motion-free k-space data as the gold standard for evaluation. 
These results of sensitivity map estimation highlight that our joint optimization strategy yields high-quality sensitivity maps for reconstruction across a range of acceleration factors.


\subsection{Results of the Motion Parameter Estimation}
Fig.~\ref{fig:motion_result} shows exemplary results of motion parameters estimation at $R = 4$ with the motion offset ($\pm 2^\circ \& \pm 4$ pixels). We compute the RMSE between the estimated motion parameters using different methods and the simulated motion states for all shots. 
The results of our proposed method exhibit lower mean values when compared to Score-MoCo (corrupted CSM), demonstrating the capability of our method for motion correction. It's noteworthy that the motion parameters predicted by JSMoCo also exhibit lower variance across all the shots. This is mainly due to the inhomogeneity of the motion-corrupted CSMs, which can lead to inaccurate motion estimations for certain shots, consequently resulting in higher variance across all the shots.

Fig.~\ref{fig:motion_t} shows the convergence behavior of our method at $R=4$ with the motion offset ($\pm 3^\circ \& \pm 3$ pixels). The model ultimately converged to generate a high-quality volume while accurately estimating motion parameters and producing reasonable CSMs within around 600 steps.

\subsection{Effects of Polynomial Representation for CSM}
Our method parametrizes the coil sensitivity map using a polynomial function, which serves a dual purpose: explicitly enforcing the continuous and smooth characteristics of coil sensitivity maps and reducing the number of unknowns to be determined, thereby mitigating the ill-posed nature of this inverse problem.
To demonstrate the effectiveness of this polynomial representation, we conducted a comparative analysis of JSMoCo with and without the polynomial representation. In the absence of polynomial representation, we parametrized the CSM as a matrix of the same size as the image and CSM was directly estimated on a pixel-wise basis.

As depicted in Fig.~\ref{fig:ablation}(b), the CSMs estimated without polynomial representation exhibit a loss of smoothness characteristics, primarily due to the absence of explicit constraints. In the polynomial representation of CSMs, the number of unknowns is proportional to $cN^2$, where $N$ is the order of the polynomial function (typically no more than 20).
However, the number of unknowns in the direct representation is $cHW$, where $H,W$ is the size of an image (typically larger than 128). 
The increase in the number of unknowns exacerbates the underdetermined nature of the inverse problem, leading to suboptimal or even erroneous estimation results.
As demonstrated in Fig.~\ref{fig:ablation}(a) and (c), the accurate estimation of CSMs plays a crucial role in both image reconstruction and motion parameter estimation.

Thus, the polynomial representation of CSMs not only reduces the number of unknowns in the optimization but also enforces a robust physical constraint throughout the MRI reconstruction process, ultimately yielding artifacts-reduced reconstruction results.

\subsection{Effects of the Accelerate Rate and Motion Levels}

Fig.~\ref{fig:model_capacity}(a) plots the variation of PSNR and SSIM of the reconstructed images as the acceleration factor increases. As the acceleration rate increases, the problem becomes increasingly underdetermined, resulting in a partial decrease in PSNR and SSIM. In Fig.~\ref{fig:model_capacity}(b) and (c), we provide PSNR and SSIM results at $R=2$ while simulating only translations or rotations, respectively. As demonstrated in Fig.~\ref{fig:model_capacity}(b) and (c), our method excels in reconstructing high-quality images within the rigid motion range ($[0,\pm 9]^\circ$ or $[0,\pm 8]$ pixels).


\section{Conclusion}
We propose JSMoCo, a novel unsupervised MRI reconstruction method that takes into account rigid motion during image acquisition. 
Different from the existing motion correction methods, our model considers the effect of motion on the coil sensitivity maps.
By incorporating the multi-coil MRI acquisition process with parameterized rigid motion and coil sensitivities into the score-based diffusion priors, we effectively 
leverage the Gibbs sampler to jointly optimize MRI images, motion parameters, and coil sensitivity maps.
Experimental results conducted on the fastMRI dataset demonstrate the superiority of our approach in reconstructing high-quality MRI images from motion-corrupted, partially-acquired k-space measurements.

\bibliographystyle{IEEEtran}
\bibliography{cas-refs}

\begin{thebibliography}{10}
\providecommand{\url}[1]{#1}
\csname url@samestyle\endcsname
\providecommand{\newblock}{\relax}
\providecommand{\bibinfo}[2]{#2}
\providecommand{\BIBentrySTDinterwordspacing}{\spaceskip=0pt\relax}
\providecommand{\BIBentryALTinterwordstretchfactor}{4}
\providecommand{\BIBentryALTinterwordspacing}{\spaceskip=\fontdimen2\font plus
\BIBentryALTinterwordstretchfactor\fontdimen3\font minus
  \fontdimen4\font\relax}
\providecommand{\BIBforeignlanguage}[2]{{%
\expandafter\ifx\csname l@#1\endcsname\relax
\typeout{** WARNING: IEEEtran.bst: No hyphenation pattern has been}%
\typeout{** loaded for the language `#1'. Using the pattern for}%
\typeout{** the default language instead.}%
\else
\language=\csname l@#1\endcsname
\fi
#2}}
\providecommand{\BIBdecl}{\relax}
\BIBdecl

\bibitem{GRAPPA}
M.~A. Griswold, P.~M. Jakob, R.~M. Heidemann, M.~Nittka, V.~Jellus, J.~Wang,
  B.~Kiefer, and A.~Haase, ``Generalized autocalibrating partially parallel
  acquisitions ({GRAPPA}),'' \emph{Magnetic Resonance in Medicine}, vol.~47,
  no.~6, pp. 1202--1210, 2002.

\bibitem{SENSE}
K.~P. Pruessmann, M.~Weiger, M.~B. Scheidegger, and P.~Boesiger, ``{SENSE}:
  Sensitivity encoding for fast {MRI},'' \emph{Magnetic Resonance in Medicine},
  vol.~42, no.~5, pp. 952--962, 1999.

\bibitem{Sparse-MRI}
M.~Lustig, D.~Donoho, and J.~M. Pauly, ``{Sparse MRI}: The application of
  compressed sensing for rapid {MR} imaging,'' \emph{Magnetic Resonance in
  Medicine}, vol.~58, no.~6, pp. 1182--1195, 2007.

\bibitem{CS-MRI-cardiac}
R.~Otazo, D.~Kim, L.~Axel, and D.~K. Sodickson, ``Combination of compressed
  sensing and parallel imaging for highly accelerated first-pass cardiac
  perfusion {MRI},'' \emph{Magnetic Resonance in Medicine}, vol.~64, no.~3, pp.
  767--776, 2010.

\bibitem{DL-MRI-MoDL}
H.~K. Aggarwal, M.~P. Mani, and M.~Jacob, ``{MoDL}: Model-based deep learning
  architecture for inverse problems,'' \emph{IEEE transactions on medical
  imaging}, vol.~38, no.~2, pp. 394--405, 2018.

\bibitem{DL-MRI-Self-super}
B.~Yaman, S.~A.~H. Hosseini, S.~Moeller, J.~Ellermann, K.~Uğurbil, and
  M.~Akçakaya, ``Self-supervised learning of physics-guided reconstruction
  neural networks without fully sampled reference data,'' \emph{Magnetic
  Resonance in Medicine}, vol.~84, no.~6, pp. 3172--3191, 2020.

\bibitem{MoCo-Review}
F.~Godenschweger, U.~Kägebein, D.~Stucht, U.~Yarach, A.~Sciarra, R.~Yakupov,
  F.~Lüsebrink, P.~Schulze, and O.~Speck, ``Motion correction in mri of the
  brain,'' \emph{Physics in Medicine \& Biology}, vol.~61, no.~5, p. R32, feb
  2016.

\bibitem{Motion-Quantify}
J.~B. Andre, B.~W. Bresnahan, M.~Mossa-Basha, M.~N. Hoff, C.~P. Smith,
  Y.~Anzai, and W.~A. Cohen, ``Toward quantifying the prevalence, severity, and
  cost associated with patient motion during clinical mr examinations,''
  \emph{Journal of the American College of Radiology}, vol.~12, no.~7, pp.
  689--695, 2015.

\bibitem{usman2020retrospective}
M.~Usman, S.~Latif, M.~Asim, B.-D. Lee, and J.~Qadir, ``Retrospective motion
  correction in multishot mri using generative adversarial network,''
  \emph{Scientific Reports}, vol.~10, no.~1, p. 4786, 2020.

\bibitem{armanious2020unsupervised}
K.~Armanious, A.~Tanwar, S.~Abdulatif, T.~K{\"u}stner, S.~Gatidis, and B.~Yang,
  ``Unsupervised adversarial correction of rigid mr motion artifacts,'' in
  \emph{2020 IEEE 17th International Symposium on Biomedical Imaging
  (ISBI)}.\hskip 1em plus 0.5em minus 0.4em\relax IEEE, 2020, pp. 1494--1498.

\bibitem{johnson2019conditional}
P.~M. Johnson and M.~Drangova, ``Conditional generative adversarial network for
  3d rigid-body motion correction in mri,'' \emph{Magnetic resonance in
  medicine}, vol.~82, no.~3, pp. 901--910, 2019.

\bibitem{8252880}
M.~W. Haskell, S.~F. Cauley, and L.~L. Wald, ``Targeted motion estimation and
  reduction (tamer): Data consistency based motion mitigation for mri using a
  reduced model joint optimization,'' \emph{IEEE Transactions on Medical
  Imaging}, vol.~37, no.~5, pp. 1253--1265, 2018.

\bibitem{cordero2018three}
L.~Cordero-Grande, E.~J. Hughes, J.~Hutter, A.~N. Price, and J.~V. Hajnal,
  ``Three-dimensional motion corrected sensitivity encoding reconstruction for
  multi-shot multi-slice mri: application to neonatal brain imaging,''
  \emph{Magnetic resonance in medicine}, vol.~79, no.~3, pp. 1365--1376, 2018.

\bibitem{batchelor2005matrix}
P.~Batchelor, D.~Atkinson, P.~Irarrazaval, D.~Hill, J.~Hajnal, and D.~Larkman,
  ``Matrix description of general motion correction applied to multishot
  images,'' \emph{Magnetic Resonance in Medicine: An Official Journal of the
  International Society for Magnetic Resonance in Medicine}, vol.~54, no.~5,
  pp. 1273--1280, 2005.

\bibitem{hossbach2023deep}
J.~Hossbach, D.~N. Splitthoff, S.~Cauley, B.~Clifford, D.~Polak, W.-C. Lo,
  H.~Meyer, and A.~Maier, ``Deep learning-based motion quantification from
  k-space for fast model-based magnetic resonance imaging motion correction,''
  \emph{Medical physics}, vol.~50, no.~4, pp. 2148--2161, 2023.

\bibitem{CCDF}
H.~Chung, B.~Sim, and J.~C. Ye, ``Come-closer-diffuse-faster: Accelerating
  conditional diffusion models for inverse problems through stochastic
  contraction,'' in \emph{Proceedings of the IEEE/CVF Conference on Computer
  Vision and Pattern Recognition}, 2022, pp. 12\,413--12\,422.

\bibitem{DPS}
H.~Chung, J.~Kim, M.~T. Mccann, M.~L. Klasky, and J.~C. Ye, ``Diffusion
  posterior sampling for general noisy inverse problems,'' in \emph{The
  Eleventh International Conference on Learning Representations}, 2022.

\bibitem{SDE-Inverse}
Y.~Song, L.~Shen, L.~Xing, and S.~Ermon, ``Solving inverse problems in medical
  imaging with score-based generative models,'' in \emph{International
  Conference on Learning Representations}, 2021.

\bibitem{Robust}
A.~Jalal, M.~Arvinte, G.~Daras, E.~Price, A.~G. Dimakis, and J.~Tamir, ``Robust
  compressed sensing mri with deep generative priors,'' \emph{Advances in
  Neural Information Processing Systems}, vol.~34, pp. 14\,938--14\,954, 2021.

\bibitem{Chung-MRI}
H.~Chung and J.~C. Ye, ``Score-based diffusion models for accelerated mri,''
  \emph{Medical image analysis}, vol.~80, p. 102479, 2022.

\bibitem{levac2023accelerated}
B.~Levac, A.~Jalal, and J.~I. Tamir, ``Accelerated motion correction for mri
  using score-based generative models,'' in \emph{2023 IEEE 20th International
  Symposium on Biomedical Imaging (ISBI)}.\hskip 1em plus 0.5em minus
  0.4em\relax IEEE, 2023, pp. 1--5.

\bibitem{Ye-motion}
G.~Oh, J.~E. Lee, and J.~C. Ye, ``Annealed score-based diffusion model for mr
  motion artifact reduction,'' \emph{arXiv preprint arXiv:2301.03027}, 2023.

\bibitem{Optim-Pattern}
S.~Ravula, B.~Levac, A.~Jalal, J.~I. Tamir, and A.~G. Dimakis, ``Optimizing
  sampling patterns for compressed sensing mri with diffusion generative
  models,'' \emph{arXiv preprint arXiv:2306.03284}, 2023.

\bibitem{peng2023one}
H.~Peng, C.~Jiang, J.~Cheng, M.~Zhang, S.~Wang, D.~Liang, and Q.~Liu,
  ``One-shot generative prior in hankel-k-space for parallel imaging
  reconstruction,'' \emph{IEEE Transactions on Medical Imaging}, 2023.

\bibitem{yu2023universal}
C.~Yu, Y.~Guan, Z.~Ke, K.~Lei, D.~Liang, and Q.~Liu, ``Universal generative
  modeling in dual domains for dynamic mri,'' \emph{NMR in Biomedicine}, p.
  e5011, 2023.

\bibitem{cui2023spirit}
Z.-X. Cui, C.~Cao, J.~Cheng, S.~Jia, H.~Zheng, D.~Liang, and Y.~Zhu,
  ``Spirit-diffusion: Self-consistency driven diffusion model for accelerated
  mri,'' \emph{arXiv preprint arXiv:2304.05060}, 2023.

\bibitem{cao2022high}
C.~Cao, Z.-X. Cui, S.~Liu, H.~Zheng, D.~Liang, and Y.~Zhu, ``High-frequency
  space diffusion models for accelerated mri,'' \emph{arXiv preprint
  arXiv:2208.05481}, 2022.

\bibitem{cui2022self}
Z.-X. Cui, C.~Cao, S.~Liu, Q.~Zhu, J.~Cheng, H.~Wang, Y.~Zhu, and D.~Liang,
  ``Self-score: Self-supervised learning on score-based models for mri
  reconstruction,'' \emph{arXiv preprint arXiv:2209.00835}, 2022.

\bibitem{Gibbs}
G.~Casella and E.~I. George, ``Explaining the gibbs sampler,'' \emph{The
  American Statistician}, vol.~46, no.~3, pp. 167--174, 1992.

\bibitem{knoll2020fastmri}
F.~Knoll, J.~Zbontar, A.~Sriram, M.~J. Muckley, M.~Bruno, A.~Defazio,
  M.~Parente, K.~J. Geras, J.~Katsnelson, H.~Chandarana \emph{et~al.},
  ``fastmri: A publicly available raw k-space and dicom dataset of knee images
  for accelerated mr image reconstruction using machine learning,''
  \emph{Radiology: Artificial Intelligence}, vol.~2, no.~1, p. e190007, 2020.

\bibitem{CS-MRI-kt-SLR}
S.~G. Lingala, Y.~Hu, E.~DiBella, and M.~Jacob, ``Accelerated dynamic {MRI}
  exploiting sparsity and low-rank structure: {kt SLR},'' \emph{IEEE
  transactions on medical imaging}, vol.~30, no.~5, pp. 1042--1054, 2011.

\bibitem{CS-MRI-Low-Rank}
K.~H. Jin, D.~Lee, and J.~C. Ye, ``A general framework for compressed sensing
  and parallel {MRI} using annihilating filter based low-rank {Hankel}
  matrix,'' \emph{IEEE Transactions on Computational Imaging}, vol.~2, no.~4,
  pp. 480--495, 2016.

\bibitem{CS-MRI-Low-Rank3}
T.~Zhang, J.~M. Pauly, and I.~R. Levesque, ``Accelerating parameter mapping
  with a locally low rank constraint,'' \emph{Magnetic Resonance in Medicine},
  vol.~73, no.~2, pp. 655--661, 2015.

\bibitem{MC-CT}
H.~Chung, B.~Sim, D.~Ryu, and J.~C. Ye, ``Improving diffusion models for
  inverse problems using manifold constraints,'' \emph{Advances in Neural
  Information Processing Systems}, vol.~35, pp. 25\,683--25\,696, 2022.

\bibitem{Adaptive-MRI}
A.~G{\"u}ng{\"o}r, S.~U. Dar, {\c{S}}.~{\"O}zt{\"u}rk, Y.~Korkmaz, H.~A. Bedel,
  G.~Elmas, M.~Ozbey, and T.~{\c{C}}ukur, ``Adaptive diffusion priors for
  accelerated mri reconstruction,'' \emph{Medical Image Analysis}, p. 102872,
  2023.

\bibitem{DSM}
P.~Vincent, ``A connection between score matching and denoising autoencoders,''
  \emph{Neural computation}, vol.~23, no.~7, pp. 1661--1674, 2011.

\bibitem{haskell2019network}
M.~W. Haskell, S.~F. Cauley, B.~Bilgic, J.~Hossbach, D.~N. Splitthoff,
  J.~Pfeuffer, K.~Setsompop, and L.~L. Wald, ``Network accelerated motion
  estimation and reduction (namer): Convolutional neural network guided
  retrospective motion correction using a separable motion model,''
  \emph{Magnetic resonance in medicine}, vol.~82, no.~4, pp. 1452--1461, 2019.

\bibitem{JSENSE}
L.~Ying and J.~Sheng, ``Joint image reconstruction and sensitivity estimation
  in sense (jsense),'' \emph{Magnetic Resonance in Medicine: An Official
  Journal of the International Society for Magnetic Resonance in Medicine},
  vol.~57, no.~6, pp. 1196--1202, 2007.

\bibitem{FSE-Motion}
B.~Levac, S.~Kumar, S.~Kardonik, and J.~I. Tamir, ``Fse compensated motion
  correction for mri using data driven methods,'' in \emph{International
  Conference on Medical Image Computing and Computer-Assisted
  Intervention}.\hskip 1em plus 0.5em minus 0.4em\relax Springer, 2022, pp.
  707--716.

\bibitem{Motion-form1}
R.~A. Zoroofi, Y.~Sato, S.~Tamura, H.~Naito, and L.~Tang, ``An improved method
  for mri artifact correction due to translational motion in the imaging
  plane,'' \emph{IEEE transactions on medical imaging}, vol.~14, no.~3, pp.
  471--479, 1995.

\bibitem{Motion-form2}
R.~A. Zoroofi, Y.~Sato, S.~Tamura, and H.~Naito, ``Mri artifact cancellation
  due to rigid motion in the imaging plane,'' \emph{IEEE transactions on
  medical imaging}, vol.~15, no.~6, pp. 768--784, 1996.

\bibitem{heckerman2000dependency}
D.~Heckerman, D.~M. Chickering, C.~Meek, R.~Rounthwaite, and C.~Kadie,
  ``Dependency networks for inference, collaborative filtering, and data
  visualization,'' \emph{Journal of Machine Learning Research}, vol.~1, no.
  Oct, pp. 49--75, 2000.

\bibitem{Langevin}
G.~E. Uhlenbeck and L.~S. Ornstein, ``On the theory of the brownian motion,''
  \emph{Physical review}, vol.~36, no.~5, p. 823, 1930.

\bibitem{NCSNv2}
Y.~Song and S.~Ermon, ``Improved techniques for training score-based generative
  models,'' \emph{Advances in neural information processing systems}, vol.~33,
  pp. 12\,438--12\,448, 2020.

\bibitem{uecker2014espirit}
M.~Uecker, P.~Lai, M.~J. Murphy, P.~Virtue, M.~Elad, J.~M. Pauly, S.~S.
  Vasanawala, and M.~Lustig, ``Espirit—an eigenvalue approach to
  autocalibrating parallel mri: where sense meets grappa,'' \emph{Magnetic
  resonance in medicine}, vol.~71, no.~3, pp. 990--1001, 2014.

\bibitem{lustig2007sparse}
M.~Lustig, D.~Donoho, and J.~M. Pauly, ``Sparse mri: The application of
  compressed sensing for rapid mr imaging,'' \emph{Magnetic Resonance in
  Medicine: An Official Journal of the International Society for Magnetic
  Resonance in Medicine}, vol.~58, no.~6, pp. 1182--1195, 2007.

\end{thebibliography}

\end{document}